\documentclass{llncs}
\usepackage{llncsdoc}

    \usepackage{main} 
    \usepackage{\filename}

    \title{\ourtitle\thanks{The final publication of this paper is available at \url{www.springerlink.com}}}
    \author{\ourselves}
    \institute{\ourinstitute\\ \ouraffiliation}

\begin{document}
    \setcounter{page}{1}
    \maketitle


    \begin{abstract}

To improve on existing models of interaction with a proof assistant (PA), 
   in particular for storage and replay of proofs, 
we introduce three related concepts, those of: 
a \demph{\prfmovie}, consisting of frames which record both user input and
the corresponding PA response;
a \demph{camera}, which films a user's interactive session with a PA as a movie; and
a \demph{\moviola}, which replays a movie frame-by-frame to a third party. 

In this paper we describe the movie data structure and we discuss a
prototype implementation of the camera and \moviola\ based on the
ProofWeb system \cite{Kaliszyk-2007}. ProofWeb uncouples
the interaction with a PA via a web-interface (the client) from the
actual PA that resides on the server. Our camera films a movie by
\scarequote{listening} to the ProofWeb communication.

The first reason for developing movies is to uncouple the reviewing of
a formal proof from the PA used to develop it: the movie concept
enables users to discuss small code fragments without the need to
install the PA or to load a whole library into it.

Other advantages include the possibility to develop a separate
\demph{commentary track} to discuss or explain the PA interaction. We
assert that a combined camera+\moviola\ provides a generic layer between
a client (user) and a server (PA). Finally we claim that movies are
the right type of \demph{data} to be stored in an encyclopedia of
formalized mathematics, based on our experience in filming the Coq standard library.


    \end{abstract}




\section{Introduction}
\sloppy 
Interaction with modern theorem-proving tools, Proof Assistants (PAs), 
   still imposes 
heavy (temporal, spatial, computational, cognitive) resource demands on users. 
It is hard to write a proof, but typically even harder to write a
formalized version 
with a 
PA. This additional overhead arises from at least 
   the following: 
\begin{itemize}
\item It is necessary to become (at least somewhat) familiar with the
  technical details of the PA, since the user needs to install and configure
     it before use; 
\item The user needs to understand the tools the PA gives her, in
  terms of libraries and commands, and how to use them to achieve
  a formalization of the proof; 
\item The user 
     needs 
  to know the proof and 
     its implicit assumptions
     in far greater detail than 
  required to communicate the 
     main ideas 
  to another person.
\end{itemize}
Much of the effort in interaction design for PAs has focused on the
second and third of these issues from the point of view of defining a
language of suitable basic proof steps, augmented with automation
layers which are either fully programmable, or else encapsulate
well-defined larger-scale proof steps, together with an editing model
of how to soundly maintain a partially completed proof.

That is to say, the basic PA use case of \realquote{writing a proof} has received most attention, while those of \realquote{reading a proof} (written by someone else) or \realquote{browsing a library} rather less so: the narrative or explanatory possibilities afforded by a formal proof text have been largely overlooked in favour of (variously prettily rendered) static digests of named definitions and theorem statements. These are necessary prerequisites for these use cases, but hardly sufficient for gaining insight into how such proofs \scarequote{work} (or even: how partial proof attempts may \emph{fail}, as when trying to discuss such examples on a mailing list). 

In each use case, however, there still remain the computational and cognitive bottlenecks arising from the first and second problems. For example, as  
illustrated by Kaliszyk \cite{Kaliszyk-2009}, before being able to
formalize a theorem in 
   Isabelle, 
a user needs to install the system, 
   comprising 
the program itself, an HOL heap and a
version of PolyML. This does not include a user interface, 
provided by the Emacs-based ProofGeneral. Having installed the PA, the
user is then left with understanding it: digging through a tutorial or
manual, and finding out what 
contributions and libraries are necessary for the formalization of the
proof.

Our work contributes to addressing the first and second issues, reducing
the overheads of installing and configuring a PA and
simplifying access to, and communication about, existing proofs. The third issue we do not address here.

Previous work at Nijmegen has partially addressed the first issue.  By
providing a generic web-interface for PAs, the ProofWeb system \cite{Kaliszyk-2009}
removes the computational load on users by uncoupling interaction with a PA via a dedicated webserver. 
This relieves the user from the
   installation problem: 
she can just visit a website and use a
web-interface to access the PA. However, when trying to understand a
(part of a) formal proof script, it is often necessary to see how the
proof state 
   changes through 
execution of a specific tactic. 
This requires first to bring the PA into that specific state 
--- finding and loading the required libraries and files --- 
   a significant overhead, not addressed by ProofWeb.

Similarly, when explaining a tactic or a part of a proof script, with
the current technology, a \prfauthor\ has to publish the \prfscript,
sometimes with an explanation of the output the PA returns to
her. This is not very satisfactory and often not informative enough,
especially when the publication of the \prfscr\ is intended to show
the intricacies of the proof: if the \prfrdr\ is uninitiated in the
specifics of the PA, she might not understand the \prfscript. 

This paper discusses the design of a further uncoupling layer,
providing a client-side Proxy to the server-side PA output. 
What we do is send a \prfscr\ piecewise to the PA and record the
response: that is, we \scarequote{film} the interaction with the
PA. So by analogy with film-editing, we have dubbed our system a
\scarequote{\moviola}
: a playback and editing suite for \scarequote{\prfmovie s} created by
submitting formal proof texts, \scarequote{scripts}, to the PA.



In the present paper we describe the basic ideas behind the
movie-camera-\moviola\ concept by discussing two use cases and a
prototype implementation based on ProofWeb, which 
can be inspected at \url{\webpage}.  

Then we further examine the versatility of the concept and observe
that we can further use our movies as a drop-in replacement for the existing
ProofWeb interaction model for editing proofs: the proof movie then becomes the \realquote{file in
the middle} that receives interactive updates from the user and the
PA.

\subsection{Contributions}


In this paper, we describe the design of an architecture for capturing
the interaction between a PA and its users. Specifically, we:
\begin{itemize}
  \item articulate two roles and associated use cases: \emph{creating} and \emph{reading} a proof; 
  \item define a \demph{\prfmovie} datastructure 
       that encapsulates interaction between a \prfauthor\ and a PA; 
    such wrapping 
       affords a \prfrdr\ fast access 
    to details of this interaction; 
  \item have built tools for creating and viewing \movie s: \demph{\camera} and \demph{\moviola};
  \item identify the set of actions an \prfaut\ needs to create and
    modify a movie on-the-fly, and the gestures that give access to
    these actions; 
  \item extend the model to allow arbitrary tools to 
       operate on movie content; and 
  \item design a concrete architecture for implementing 
       such a system. 
\end{itemize}



\section{Background and Use Cases}
\label{sec:background}
We identify two roles involved in communicating a \prfscript: the
\demph{\prfauthor} and a \demph{\prfreader} (the reader).  The
\prfauthor\ is a user of a PA that creates a (formal) proof. The
interaction with the PA is encoded in a \demph{\prfscript}, which
contains the commands used to build the proof.
The author can communicate the proof to any \prfrdr\ by publishing
this \prfscr, and the \prfrdr\ can look at the \prfscr\ by loading it
in his local version of the PA.

\begin{figure}
  \begin{center}
    \includegraphics[scale=.3]{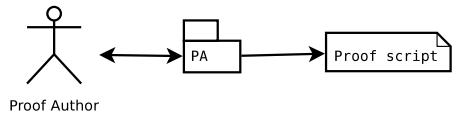}
  \end{center}
  \caption{Creating a proof script \label{fig:UCcreate}}
\end{figure}

\begin{figure}
  \begin{center}
    \includegraphics[scale=.3]{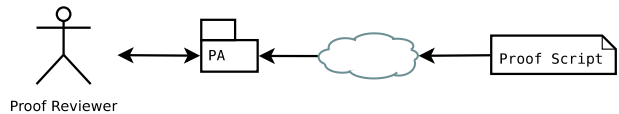}
  \end{center}
  \caption{Reading a proof script \label{fig:UCread}}
\end{figure}

These two roles both have their own well-defined activities, but an
actor (an actual user of a system, instead of just an abstract role)
can play both the author-role and the \prfrdr-role: when writing a
proof, an author might want to review what he has done before.
To make the activities of both roles explicit, we identify a single
use case for each of them. These use cases are \emph{creating a proof} for
the \prfaut\  
(Figure~\ref{fig:UCcreate}), 
and \emph{reading a proof} for the \prfrdr\  
(Figure~\ref{fig:UCread}).

A third use case is \emph{browsing a library}, in which a user takes on a role similar to that of the \prfrdr, but also searches for useful lemmas in the library, and tries to understand how they are used. While we do not treat this use case here (we are grateful to one of our referees for drawing attention to it), we nevertheless claim that such \scarequote{advanced} elaborations of the reviewer's use case would benefit from the \moviola\ technology presented here. Current technology does not give users fast access to other developments, which is a prerequisite for this use case.

   \paragraph{Legend} 
The diagrams  
   below are 
almost, but not quite, 
    standard UML: 
\begin{itemize}
\item A stick figure represent a user role (\prfauthor\ and \prfreader).
\item A \scarequote{package} represents a tool/program instance (here: the PAs). 
\item The cloud represents the Internet.
\item A folded page is a file (here: the \prfscript).
\item Arrows represent data flow; a double arrow,  
  interaction between two parts.
\end{itemize}

\paragraph{First Use Case: Creating a proof}
To create a proof, 
   an \prfaut\ writes commands to be interpreted
by the PA. 
   In response, 
the state of a proof changes by
   decomposing 
the theorem to be proved, generating new proof
obligations or 
   discharging 
goals as proven. 
   A \prfscript\ stores a a transcript of the commands issued. 

In Figure~\ref{fig:UCcreate}, we display the traditional implementation of this use case, in which the PA is locally installed, and creates a local copy of the \prfscript.


\paragraph{Second Use Case: Reading a proof}
To read a proof, the \prfrdr\ obtains a (copy of a)
\prfscript, possibly via the Internet. 
   Subsequently, 
he can load it in his copy of the PA,
and \scarequote{replay} the proof: many 
   PA interfaces 
have a notion of
stepping through a \prfscr, by sending commands one-by-one to
the PA or by undoing the last command sent.  Because the PA does not
know that the commands it receives are extracted from a \prfscr, it
responds to commands in the same way as in the creation use case: by
sending the new proof state. 


\paragraph{Discussion} This form of communication has a number of problems:
\begin{enumerate}
\item If only a small part of the \prfscr\ is relevant, it might still be necessary to send the entire \prfscr\ to the \prfrdr: the parts of interest might require definitions defined previously in a \prfscr, or might use lemmas proved earlier. 
\item Before a proof can be reviewed, the \prfrdr\ needs to install (ideally the same version of) the PA the \prfaut\ used, or be so familiar with 
   its 
technology to interpret the \prfscr\ mentally\footnote{For a PA that uses a procedural proof style (tactics), 
    it is hopeless to try to interpret a proof script purely mentally, 
    without seeing it executed.
}. Especially when the \prfscr\ is used to communicate a proof to a \prfrdr\ who is not part of the PA community, this can be a large handicap.
\end{enumerate}

\noindent A possible solution to the 
   first problem, 
frequently exercised on the Coq-club mailing list \cite{CoqClub}, is to simplify a \prfscript\ to a minimal example, which focuses on the problem in the \prfscr, and the definitions directly necessary to obtain this problem. 
   But such 
simplification might be too 
   drastic, 
abstracting away crucial details. Furthermore, abstraction is not an option when the main purpose of the communication is not to point out a problem, but to display a (partial) formalization of a proof to an outsider, because it is necessary to stay close to the vocabulary and methods of the target audience.

The ProofWeb system developed by Kaliszyk \cite{Kaliszyk-2007} places the PA on a central server that is accessible through an AJAX-based web application. This means that to review a proof, a \prfrdr\ needs only a web browser and the \prfscript, which could be hosted on the same server as the PA.

  ProofWeb is an Internet-mediated realisation of the \scarequote{creation} use case, which mitigates the 
   second problem 
of having to install and configure a PA, but does not yet allow partial communication of the proof to a \prfrdr. It does not provide fast access to an arbitrary proof state: to obtain the state after a given command, all preceding ones need to be resent to the PA for reprocessing.




\section{The \Prfmovie}

In the previous section, we identified two problems with the current method of distributing a \prfscript\ from a \prfauthor\ to a \prfreader. In summary, these problems are that reviewing requires the \prfreader\ to resubmit the \prfscript\ to the PA, and that the \prfscript\ cannot be reviewed in fragments.

To solve these problems, we propose to enrich the \prfscript\ data structure. In the new data structure, which we call the \demph{\prfmovie}, we record the commands sent to the PA coupled together with the response of the PA to each command. Such a pair of a command and a response is a \demph{frame}. The exact Document Type Definition for the \movie\ data structure can be found at \url{\webpage/movies/film.dtd}. 


   The \prfmovie\ 
   is designed 
to be \emph{self-contained} and \emph{generic}
:
\begin{description}
\item[Self-contained] Making the \movie\ self-contained means that a \prfreader\ only needs a \movie\ and a tool capable of displaying it to replay the proof: no other tools are necessary for this. Aside from this, the frames themselves contain the exact state of the proof at the point represented, meaning that it is possible for an \prfaut\ to publish a proof partially, omitting or reordering frames before publication.
\item[Generic] By making the \movie\ generic, 
    creation and display do       
not depend on a specific PA or PA version: this does require that 
    one can 
specify a transformation from the PA's interaction model 
    generating discrete frames. 
\end{description}

\begin{figure}
\small
\begin{center}
\begin{tabular}{|l|}
\hline
  \begin{minipage}{.95 \textwidth}
\begin{alltt}
\strut<frame frameNumber="2">
  <command>
    intros A x.
  </command>
  <response>
    1 subgoal
  
    A : Type
    x : A
    ============================
    A
  </response>
\strut</frame>
\end{alltt}
\end{minipage} \\
\hline
\end{tabular}
\normalsize
\end{center}
\caption{An example frame of a Coq movie \label{fig:frame}}
\end{figure}


To implement the \movie\ data structure, we used the eXtensible Markup
Language (XML). 
An example 
frame in this
implementation can be found in Figure~\ref{fig:frame}.
This example contains the notion of a \scarequote{frame number}
: a sequence number 
identifying the order in which the commands were submitted to the PA.

We do not consider the \movie\ to be a complete replacement for a \prfscr. Instead, it is a container of a part of the \prfscr, together with the output of the PA. This output does not need to be correct, but this does not interfere with our intention of the \movie: we see a \movie\ as an explanation of a proof, not as checked \prfscript\ \perse. If a \movie\ contains a complete \prfscr, the concatenation of all the command segments of all frames in the \movie\ reproduces the \prfscr, which can then be (re-)checked by a PA.

The \movie\ introduces a new use case, \emph{creating a \movie}, which we describe in Section~\ref{sec:construct}. Having the \movie\ also changes the use case of reviewing a  \prfscript, and we describe this 
   modified 
use case first, in Section~\ref{sec:watching}.



\section{The Proviola: Watching a \Prfmovie}
\label{sec:watching}

When the \prfrdr\ has obtained a \prfmovie, he wants to access the data within it to review the proof, much like when he obtained a \prfscript. 
In effect, the \realquote{reading a \prfscr} use case described in Section~\ref{sec:background} and illustrated in Figure~\ref{fig:UCread} has not changed, only the data structure 
   supporting 
   it 
has changed.

We call the system used for displaying a \movie\ a \demph{\moviola}.
Just as in film-making, where an editor might wish to review a film while editing, and moreover quickly fast-forward and rewind the \movie\ to see individual shots, we wish to achieve similar access speed and portability. 
Indeed, our \moviola\ is not a separate tool: rather, we realize it through the use of HTML and very simple JavaScript.

\paragraph{Reading a proof script, revisited: watching the movie}
\begin{figure}
  \begin{center}
    \includegraphics[scale=.3]{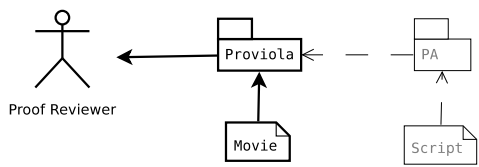}
  \end{center}
  \caption{Watching a movie: \moviola\ and \movie\ proxy PA behaviour \label{fig:UCreadMovie}}
\end{figure}

The \movie\ is self-contained, and can be distributed like a \prfscr. Unlike a \prfscr, the contents of a \movie\ can be inspected without any external tools except a web browser: the \movie\ can be located anywhere, and inspected from this location. In particular, a PA is not required to compute the proof's state and the \movie\ can be watched offline, at any time.
After the \prfrdr\ loads a \movie\ in the \moviola, he wants to step through the proof much like when a PA was loaded with a \prfscr: by indicating for which command he wants to see the response. 

As illustrated in Figure~\ref{fig:UCreadMovie}, the \moviola\ and the \movie\ together \demph{proxy} \cite[Chapter 5]{GangOfFour-1994} the behaviour of a PA: the responses shown to the \prfreader\ are stored (or cached) in the \movie, after having been computed by the PA. 
\paragraph{A prototype implementation of the \moviola} 

\begin{figure}[!htp]
\begin{center}
\includegraphics[width=.95 \textwidth]{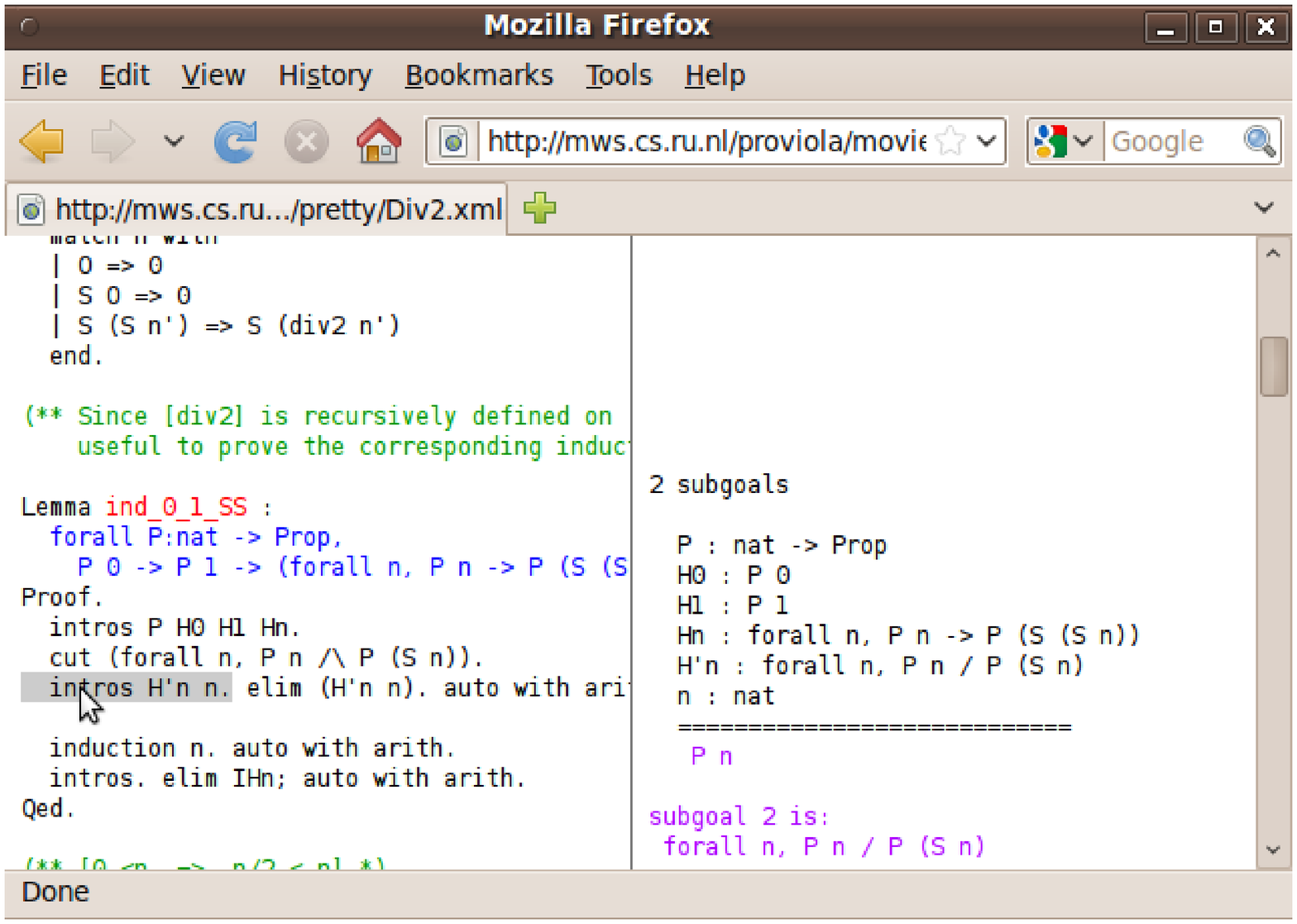}
\end{center}
\caption{Screenshot of a \moviola \label{fig:screenshot}}
\end{figure}

We implemented a prototype of the \moviola\ as an XSLT transformation from the \movie\ into an HTML file containing embedded JavaScript. This page initially shows the commands within the \movie, much like a \prfscript.  
Figure~\ref{fig:screenshot} illustrates this: 
    when 
the \prfrdr\ places his cursor over a command, the corresponding response is revealed  
   dynamically.
   The command pointed to is highlighted as a visual reminder. The prototype \moviola\ can be inspected at \url{\webpage/movies/movies.html}.








\section{The Camera: Creating a \Prfmovie}
\label{sec:construct}
\begin{figure}
  \begin{center}
    \includegraphics[scale=.3]{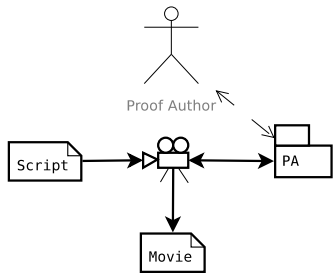}
  \end{center}
  \caption{Creating a movie: camera and script proxy user behaviour \label{fig:UCwriteMovie}}
\end{figure}

   Before a  \movie\ can be replayed, 
it 
   must be 
   created 
from a \prfscript\ by a tool 
we call a \demph{\camera}. 
   Such 
\emph{creation of a \movie} is a new use case, 
   shown 
in Figure~\ref{fig:UCwriteMovie}.
\paragraph{Third Use Case: Creating a \movie}

   This 
is non-interactive: the user of the \camera\ invokes it on a \prfscr, after which the tool does all the work, 
   yielding 
a \movie.

After it has been invoked, the \camera\ parses the \prfscript\ given to it into separate commands. These commands are stored in a frame and sent to the PA. When the PA responds to a command, the response is recorded alongside the command. The frame is subsequently appended to the \movie.

From the perspective of the PA, the \camera\ and the \prfscr\ behave like an actual \prfauthor. In other words: in creating the \movie, the \camera\ and \prfscr\ together \emph{proxy} the behaviour of a \prfauthor.


Because the \prfaut\ holds the original \prfscr, it seems natural that she invokes the \camera\ on it to obtain a \movie, 
   for distribution 
to \prfrdr s. However, a \prfrdr\ might also 
   play the role 
of cameraman, given access to the \prfscr. 

   \paragraph{Prototype implementation}
We implemented the \camera\ as a client to the ProofWeb 
   system, available 
at \url{\webpage/camera/camera.html}. 
Although it is possible for the \camera\ to communicate directly with each PA, we believe that using a generic wrapper like ProofWeb has the following advantage:
ProofWeb provides a generic interface to different PAs: 
   the communication protocol 
is the same for each PA, and the only
PA-specific knowledge the \camera\ needs to have is how to split a
\prfscript\ into separate commands.  

The main disadvantage to this approach is that to support new PAs with the \camera, these need to be made compatible with ProofWeb, which imposes stricter requirements on the interaction model than the \movie\ design needs.
In our design, the \camera\ behaves as a straightforward client to a ProofWeb server, wrapping the commands in the annotation expected by that server and unwrapping the responses.
A step beyond this, which we investigate in Section~\ref{sec:proxy}, is creating a \movie\ automatically and on-demand.



\section{Proxying Movie-Making}
\label{sec:proxy}


Until now, we have created \prfmovie s by submitting a complete script to a PA. In particular, we have filmed the Coq standard library \cite{Coq-stdlib}. We now wish to go beyond this simple case of filming completed scripts, and investigate how to create a \movie\ dynamically, by observing the interaction between \prfauthor\ and the PA. Based on these observations, we redesign this architecture to support the desired behaviour. This architecture will be implemented in future work. 

\begin{figure}
  \begin{center}
    \includegraphics[scale=.3]{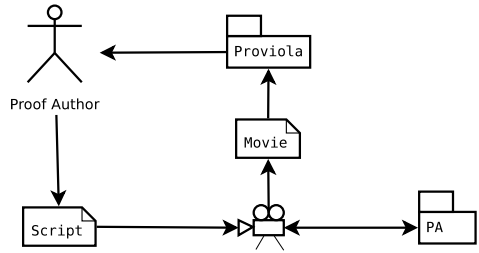}
  \end{center}
  \caption{Interactively creating a movie: instantiating the proxies \label{fig:UCinteractiveMovie}}
\end{figure}

As we have mentioned in Section~\ref{sec:background}, a user taking the role of a \prfauthor\ can also take on that of \prfreader. If she takes on these two roles simultaneously, we get the situation depicted in Figure~\ref{fig:UCinteractiveMovie}. This figure is constructed by composing Figures~\ref{fig:UCreadMovie} and~\ref{fig:UCwriteMovie}, replacing the proxied components from each figure with their implemented counterpart in the other figure.

In this figure, the \prfauthor\ writes commands into a \prfscr, which is submitted to the camera command-by-command. The command is then handled by the camera as described in Section~\ref{sec:construct}. Because the \prfauthor\ has the \movie\ corresponding to the \prfscr\ loaded in the \moviola, it updates whenever the script updates or when the PA responds to a command. 

The camera as designed in the previous section requires an explicit action by the creator of the movie: the camera is a tool that needs to be invoked on a \prfscript\ to create a movie from it. Such a design requires the \prfauthor\ to constantly update the movie when updating the  \prfscr, to keep them in sync. In fact, if we follow the information flow in Figure~\ref{fig:UCinteractiveMovie}, the \prfaut\ cannot even see the result of her own changes to the \prfscr\ if she does not use the \camera\ first.

As a solution for this disadvantage we merge the \prfscr\ and \movie\ into a single data structure, that is manipulated by both the \prfauthor\ (through the \moviola) and the PA. To obtain this, we will need to modify the \movie, the \camera\ and the \moviola\ in the following ways:

\paragraph{Movie} The \movie\ does not need to change in any radical way. The only change 
   necessary is that a \movie\ be editable  
after it has been created. This way, the \prfauthor\ can write commands into the \movie\ as she would do into the \prfscr.

\paragraph{Proviola} The \moviola\ already provides a display of the \movie, giving the \prfreader\ access to the \prfscr\ and the proof state at that point. To allow an \prfaut\ to update the \movie, the \moviola\ needs to be extended with an interface to update the commands in the frames in the \movie. 
This extension is done by adding the notion of a \demph{focused frame}, which can currently be edited. To manipulate this frame, we add gestures for the following actions to the \moviola: 
\begin{description}
  \item[Create a movie] This action creates a new, empty movie.
  \item[Focus frame] The \prfauthor\ can use this action to change the focused frame to be the frame she is interested in editing.
  \item[Edit frame] The only frame that can be edited is the focused frame.  
  \item[Add frame to a movie] When the \prfaut\ finishes a command in the focused frame, a new frame is tacitly added to the \movie\ and given focus. The previously focused frame is submitted to the camera for further processing.
  \item[Remove frame from a movie] This is an inverse action to adding a frame.
\end{description}
We consider submitting a frame for further processing an implicit action: the \prfauthor\ does not indicate in any way that she is finished with a frame, but the system recognizes a frame to be complete and processes it further, including rechecking commands later in the \movie, if these depend on the frame that was changed. What \scarequote{depends on} actually means depends on the script-management model (and more generally: the interaction model) of the PA used.

\paragraph{Camera} To keep the \moviola\ as lightweight as possible, the PA's manipulation of the \movie\ is brokered through the \camera. This includes periodically checking  whether 
   the \prfaut\ completed a command 
in the focused frame. Frames containing completed commands are then split off the focused frame and submitted to the PA. 
   This means the \camera\ evolves from batch-processing a \prfscr\ (as in Section~\ref{sec:construct}) to continuously reading the \movie, updating its contents as needed.
%
%

\begin{figure}
  \begin{center}
    \includegraphics[scale=.3]{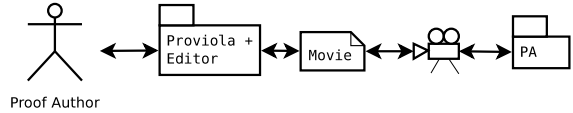}
  \end{center}
  \caption{Interactively creating a movie: cutting out the script \label{fig:UCinteractiveMovie-final}}
\end{figure}

By merging \prfscr\ and \movie, implementing the changes above, we obtain the architecture shown in Figure~\ref{fig:UCinteractiveMovie-final}.

\paragraph{Design and implementation of the system}
We have embarked upon a prototype implementation of the
architecture described above. 
We decided to 
base it on ProofWeb's architecture, 
placing the \camera\ on a server, with the \moviola\ as a client to the \camera, as an interface to the services offered at the server. 

The \movie\ is kept both as a local copy by the \moviola\ and as a remote copy by the \camera, and the protocol between \camera\ and \moviola\ is meant to keep these copies synchronized: each action of the \prfaut\ is 
   executed 
on the local copy and then communicated to the \camera, which communicates the change to the PA. As such, the \movie\ becomes a \emph{proxy} for the PA's state, and the interaction of \prfauthor\ with PA is proxied by editing the \movie.

By being placed between client and PA, the camera  
plays a similar role as the Broker in the PG Kit framework \cite{AspinallLuethWinterstein-2007}. The main differences between that system and our \moviola-based system, are based on the design decision to make the movie the main entity of the architecture. This has the following implications:
\begin{description}
\item[Cached history] The history of the interaction with the PA is stored as a movie. Because the client has a local copy of the movie, it provides instant access to the history of the proof development.
\item[Implicit proof navigation] The \prfaut\ can freely choose the frames of the movie to edit instead of having to request the PA to unlock or process a line.  
\end{description}

To synchronise the \movie\ between \camera\ and \moviola, we follow a version of the \emph{publisher/subscriber} \cite[Chapter 4]{GangOfFour-1994} design pattern:
\begin{itemize}
\item The \moviola\ holds the local copy of the \movie, and records when a user focuses on a specific frame.
\item When a frame's content changes, for example, when a PA responds to a command, the \camera\ notifies the \moviola\ of this.
\item If this frame has not been loaded before, or has been updated, it is requested from the \camera, and cached locally.
\item If the frame has been requested before, and has not been updated, it is loaded from cache.
\item When the user updates the focused frame, it is sent to the \camera. 
\end{itemize}




\section{Applications}

\subsection{The Camera as a Service Broker}

In our design of interactive \movie-making, the \camera\ takes a centralized place at the server, containing the \scarequote{master copy} of the \movie\ and providing write access to it. This access is not only usable by the \moviola\ or the PA, but could also be used by other tools that work with a formal proof. 

To provide access to the \movie, the \camera\ needs to allow arbitrary tools to register as either a \emph{subscriber} or as a \emph{publisher} for a \movie.

\paragraph{Subscribers} A \demph{subscriber} to a \movie\ can obtain the frames of the \movie\ in which they are interested. When a frame is changed in the \camera-side \movie, it notifies each subscriber of the changed frame. Subscribers can then update the frame.
   Our intention is that subscriber access be read-only. 

\paragraph{Publishers} 
   By contrast, when 
registered as a \demph{publisher}, a tool can write in the \movie. We do not believe there should be any restriction to the types of content that publishers can write, but do require that the \movie\ should be \emph{extended}: if the tool produces information of a different type than already exists in the \movie, it should be added alongside the existing data, not replace it. So, although it is possible for multiple tools to write into the command section of a frame, for example a tool processing a \prfscr\ and the \moviola+editor of Section~\ref{sec:proxy}, a tool that does not produce commands should not write in the command section.

The idea of tools listening for changes in 
proof state is not new: it was previously mentioned by Aspinall, L\"uth and Winterstein \cite{AspinallLuethWinterstein-2007} as future work for their broker architecture. Using the \camera\ as a broker is 
   similar,  
with the advantage that history does not need to be kept by the individual tools but can be requested from the \camera: an additional benefit is that subscribers coming late to the party still have access to the full history of a proof, 
   which we consider 
especially interesting for our 
   ongoing 
work towards a repository of formal proofs. 

\subsection{Towards an Encyclopedia of Formal Proof}
With some modifications, the \prfmovie\ can be used as the data structure
underlying an encyclopedia 
that
we envisage 
containing
formal proofs together with an informal narrative explanation, and provide a
toolbox for using and manipulating 
   such composite \scarequote{articles}, 
as originally sketched in  \cite{p3}.  

The \movie\ provides a generic structural view of different kinds of
   \prfscr, so that 
for a tool wanting to manipulate a proof on a
structural level, it is not necessary to know the exact details of the PA 
   concerned, 
and the history contained in the
\movie\ implies that tools 
   developed later 
are not required to replay the entire \prfscr\ through a PA.
Furthermore, because the \camera\ provides a central interface for
accessing the \movie, tools do not have to be on the server hosting
the encyclopedia, allowing others to use the proofs without 
   undue 
restrictions.

To implement the encyclopedia, we do need to create several tools
ourselves, that together form the backbone of the encyclopedia.

First, we need to store and retrieve the \movie. Possible candidates for storage are
the file system, a database or a version control system which keeps
track of the history of a proof. Retrieval cannot easily be expressed
by communicating with the \camera, and will most likely warrant an
extension to that concept.

The second addition is aimed at adding informal explanation to a
proof: a \scarequote{commentary track}, where an author can comment on the
proof script and the output that the PA produces.
This requires a revision of the \movie-concept. Because several frames
in a \movie\ might be explained in a single, continuous narrative and
commands might be repeated several times in the narrative, we will add
\emph{scenes} to the \movie.
A \demph{scene} contains a single, informal description of a group of frames,
and a reference to these frames. A scene can refer to zero or more
frames, and a frame may belong to zero or more scenes.

As well as linear narrative, scenes could be used to store alternative problem-solving approaches, such as using different lemmas or automated proof search. 

As previously mentioned, we have investigated the \movie's ability to capture a large body of proofs by filming the Coq standard library \cite{Coq-stdlib}. The
filming took less than one hour, including a two-second sleep between
processing each file, which was inserted to prevent overloading the
server running ProofWeb.
The resulting films can be inspected at
\url{\webpage}, which serves the films quickly,
even for large developments such as that 
   of the Riemann integral, at 
around \megabytes[two]\ in size. Movies typically are much smaller, however, up to 
\kilobytes[five hundred]. The \movie\ size scales in the size of the original script: a long script that uses tactics producing many subgoals also creates a large \movie.

In further work \cite{TankinkGeuversMcKinna-2010}, we elaborate the concepts of scenes and
commentary, and 
describe tooling for creation and
display of a commented \movie, by investigating \movie s of course notes on Software Foundations by Pierce \etal \cite{Pierce+-SF}, available at \url{\webpage/movies/sf}.




\section{Related work}
\newcommand{\tmegg}{tmEgg}
\newcommand{\texmacs}{\TeX macs}



We have already mentioned 
   the PG kit system and its
protocol, PGIP \cite{AspinallLuethWinterstein-2007}. 
   Interactive \movie-making 
is similar to 
   its broker architecture. 
While PGIP focuses on the message structure for 
   the dynamics of communication between \prfaut\ and PA, 
a \movie\ freezes 
   a sequence of such messages 
into a 
   static 
data structure. In that respect,
   our 
   approach 
is orthogonal to 
   that of PGIP: 
the \movie\ stores the result of an interactive session while PGIP deals
with the messages and protocol 
needed to generate it. 
   Instead of a prototype based on the \camera\ and ProofWeb, we could 
   just as well 
have filmed an interaction with the PG kit broker.

Wenzel's Jedit editor \cite{Wenzel-2009} also considers a finer
grained analysis between editor and Isabelle. But this is more
oriented towards 
   parallelisation of proof steps than the 
human-orientated details of interaction and representation of proofs 
   considered here. 

The \movie\ file format is an XML-based representation of formal
mathematical documents.
The best-known and most versatile XML-based format for mathematical
documents is OMDoc \cite{koh:06}.
We decided not to use this format internally in our system.
OMDoc is about document \emph{structure}, while
the content of a movie is unstructured system input/output.
Still, if needed, movies could be easily converted to OMDoc, so our choice of a simpler format is not a limitation.


The history stored in a \movie\ is not the same as history in, say, a
web browser: the \movie\ stores both the messages sent and the
responses received, while a browser only stores a reference to pages
visited, and needs to obtain the page again when the user requests it.
This approach of \scarequote{recalculating} a result is similar to what happens
when a user sends undo commands, both when using a normal program
(such as a text editor) and when using a PA. In our \moviola\ there is
no recalculation, because all the \scarequote{history}  
is stored in the \movie. 

A command in a PA can consist of a combination of
more primitive tactics, for example two tactics sequentially composed
with the ``;'' tactical in Coq. One might wish to see the
intermediate state after the first tactic in such a sequence. Coq does
not support this small step execution model, so in our implementation,
the movie cannot provide this information. A system like Matita
\cite{matita} does, so a \moviola\ based on Matita could potentially
expose this refined execution.

Integrating a \scarequote{commentary track} into the movie, by
collecting frames into \scarequote{scenes} and adding a narrative text to it,
goes in the direction of earlier work on
\tmegg\ \cite{Geuvers-Mamane-2006}. There we 
   started with 
a mathematical document, written in the editor
\texmacs, as the backbone 
   with 
a Coq proof script (a \texttt{.v} file)
underneath it. At any point 
one could consult the
formal proof by opening a Coq 
   session 
within the document and
executing Coq commands. This means Coq is started up and brought
into the required state, and then the selected commands are
executed. This works quite well for small \prfscr s, but can take
a lot of time for larger proofs. Also, it requires the PA to be
available, with the right libraries, which makes the mathematical
document less \scarequote{self-contained}. Another hindrance is that
\tmegg\ relies heavily on \texmacs, which means that one is
dependent on yet another application to run smoothly. The present work
allows \movie s to play the role of the Coq interaction (as a
proxy) within an interactive mathematical document. This can basically
be done within any editor and thus 
   relieves 
the interactive
mathematical document from being tied to either \texmacs\ or Coq.



\section{Conclusion\label{sec:conclusions}}

In conclusion, we claim that our refactored interaction model and its
associated data structure are an important contribution in their own
right. But our interest is in how this data structure may be further
extended to support richer interaction and display as part of a
MathWiki.


\paragraph{Acknowledgments}
  The idea of \movie-making emerged from discussions with Dan Synek about the absence of a lightweight way to communicate a formal proof to others. 
We would like to thank the anonymous referees, whose questions and suggestions helped improve this paper.
This research is partially funded by the NWO project \realquote{MathWiki: a Web-based Collaborative Authoring Environment for Formal Proofs}, the NWO BRICKS/FOCUS project \realquote{ARPA: Advancing the Real use of Proof Assistants} and the NWO cluster \textsc{Diamant}.

\bibliographystyle{splncs03}
\bibliography{\filename}

\end{document}